\documentstyle[12pt]{article}
\def\hybrid{\topmargin -20pt	\oddsidemargin 0pt
	\headheight 0pt	\headsep 0pt
	\textwidth 6.25in	% A4 paper
	\textheight 9.5in	% A4 paper
	\marginparwidth .875in
	\parskip 5pt plus 1pt	\jot = 1.5ex}

\hybrid

\begin{document}
%\titlepage
\def\x{\times}
\def\ra{\rightarrow}
\def\beq{\begin{equation}}
\def\eeq{\end{equation}}
\def\beqa{\begin{eqnarray}}
\def\eeqa{\end{eqnarray}}
\def\D{ {\cal D}}
\def\L{ {\cal L}}
\def\C{ {\cal C}}
\def\N{ {\cal N}}
\def\calE{{\cal E}}
\def\lin{{\rm lin}}
\def\Tr{{\rm Tr}}
\def\mxth{\mathsurround=0pt }
\def\xversim#1#2{\lower2.pt\vbox{\baselineskip0pt \lineskip-.5pt
x  \ialign{$\mxth#1\hfil##\hfil$\crcr#2\crcr\sim\crcr}}}
\def\simgr{\mathrel{\mathpalette\xversim >}}
\def\simle{\mathrel{\mathpalette\xversim <}}

\def\a{\alpha}
\def\b{\beta}
\def\dota{ {\dot{\alpha}} }
\def\lag{Lagrangian}
\def\Kahler{K\"{a}hler}
\def\kahler{K\"{a}hler}
\def\A{ {\cal A}}
\def\C{ {\cal C}}
\def\D{ {\cal D}}
\def\F{{\cal F}}
\def\L{ {\cal L}}
\def\R{ {\cal R}}
\def\x{ \times }
\def\beq{\begin{equation}}
\def\eeq{\end{equation}}
\def\beqa{\begin{eqnarray}}
\def\eeqa{\end{eqnarray}}

%\addtocounter{section}{1}

\renewcommand{\thesection}{\arabic{section}.}
%\renewcommand{\theequation}{\thesection \arabic{equation}}

%\setcounter{section}{1}
%\addtocounter{section}{1}

\parindent0em

\def\beq{\begin{equation}}
\def\eeq{\end{equation}}
\def\beqa{\begin{eqnarray}}
\def\eeqa{\end{eqnarray}}

\sloppy
\newcommand{\be}{\begin{equation}}
\newcommand{\eq}{\end{equation}}

%\textwidth14.5cm
%\textheight23.0cm
%\oddsidemargin0.5cm
%\topmargin-1.4cm

%\addtocounter{section}{1}

\begin{titlepage}
\begin{center}

\hfill CERN-TH/96-70\\
\hfill HUB-EP-96/7\\
\hfill hep-th/9603108\\

\vskip .8in

{\large \bf Instanton 
Numbers and Exchange Symmetries in $N=2$ Dual String Pairs}

\vskip .2in

{\bf Gabriel Lopes Cardoso$^a$, Gottfried Curio$^b$,
Dieter L\"ust$^b$ and Thomas Mohaupt$^b$}\footnote{email: 
cardoso@surya15.cern.ch,
	curio@qft2.physik.hu-berlin.de,
	luest@qft1.physik.hu-berlin.de,\\
	mohaupt@qft2.physik.hu-berlin.de}
\\
\vskip 1.2cm
$^a${\em Theory Division, CERN, CH-1211 Geneva 23, Switzerland}

$^b${\em Humboldt-Universit\"at zu Berlin,
Institut f\"ur Physik\\
D-10115 Berlin, Germany}

\vskip .1in

\end{center}

\vskip .2in

\begin{center} {\bf ABSTRACT } \end{center}
\begin{quotation}\noindent
In this note, we comment on Calabi-Yau spaces with Hodge
numbers $h_{1,1}=3$ and $h_{2,1}=243$.
We focus on the Calabi-Yau space $WP_{1,1,2,8,12}(24)$
and show how some of its instanton numbers are related
to coefficients of certain modular forms.  We also comment
on 
the relation of four dimensional exchange symmetries
in certain $N=2$ dual models
to six dimensional heterotic/heterotic string duality.

\end{quotation}
\vskip 4cm
CERN-TH/96-70\\
March 1996\\
\end{titlepage}
\vfill
\eject

\newpage

\section{Introduction}

Recently, there has been an enormous progress in the understanding
of non-perturbative effects in supersymmetric field theories 
and in superstring 
theories. In particular, various types of strong-weak coupling
duality symmetries
are by now quite well established, such as $S$-duality 
 of the 
four-dimensional
$N=4$ heterotic string \cite{sduality,SchSen,Sen2}, 
string/string dualities \cite{HullTown,Wit1,KaVa} 
between the heterotic and type 
II strings 
and heterotic/heterotic duality in $D=6$ \cite{DuMinWit,AlFoIbaQue2}. 
It now seems that most or even all  non-perturbative duality
symmetries originate from underlying theories in higher dimensions
(from $M$-theory \cite{Wit1,Schwarz} in
 $D=11$ or from $F$-theory \cite{Vafa,MoVa} in $D=12$).

In the following we will be concerned with $N=2, D=4$ string theories
which have both a heterotic and a type II description \cite{KaVa,KLM}.
%in the sense
%of strong-weak coupling string/string duality. 
In this context, there 
is a particularly interesting class of models, which exhibits
a non-perturbative symmetry which exchanges the heterotic dilaton $S$
with one of the vector moduli fields. 
In the context of heterotic/type II string/string duality this type
of exchange symmetry was first discussed in \cite{KLM},
where this exchange symmetry was related to a 
a remarkle symmetry property of rational instanton numbers. 
Subsequent work 
connected this symmetry to the monodromy group of the CY 
compactification  
\cite{KKLMV,AntoPartou} and discussed \cite{CLM2} 
the action of this exchange symmetry on BPS
spectra and higher derivative gravitational couplings.
In this paper, we will extend the previous work of
\cite{KaVa,KLM,KKLMV,KaLouThei,Curio1,Curio2} in several directions.
We will focus on a class of type II theories based on
elliptically fibered CY spaces.  We will, in particular,
discuss the CY space described by a hypersurface of degree 24
in
weighted projective space $WP_{1,1,2,8,12}(24)$ 
with, in heterotic language, three vector moduli $S,T,U$.
We will investigate the rational (genus 0) as well as the elliptic (genus 1)
instanton numbers for this class of models. 
We will show  that the genus 0 as well as
the genus 1 instanton numbers can, 
in the heterotic weak coupling limit, be  precisely expressed by
 the coefficients
of the $q$ expansion of certain modular forms. This means that these instanton
numbers 
are nothing else than the 
multiplicities of positive roots
of some generalized Kac-Moody algebra recently discussed in \cite{HarvMoo}.
(For the rational instanton numbers this relation was already anticipated in
\cite{HarvMoo}.) Hence one can expect that the non-perturbative 
$S-T$ exchange symmetries are reflected in a nice symmetry structure of some
non-perturbative infinite-dimensional algebra.
In order to relate these instanton numbers to the expansion coefficients
of certain modular forms, we will have to
work out the precise identification of the heterotic vector moduli
with the corresponding type II 
K\"ahler class fields.  We will also
discuss the action of the $S-T$
exchange symmetry in this context. 
At the end, we will comment on the relation of the
four-dimensional exchange symmetry to the six-dimensional heterotic/heterotic
duality symmetry \cite{Duff,DuMinWit,AlFoIbaQue2}
in this class of models.

\section{Instanton numbers and exchange symmetries}

The higher derivative couplings of vector multiplets $X$
to the Weyl multiplet ${\cal W}$ of conformal $N=2$
supergravity can be expressed as a power series \cite{AnGaNarTay,confsug}
\beqa
F(X,{\cal W}^2)=\sum_{g=0}^\infty F_{g}(X)({\cal W}^2 )^g.\label{expansion}
\eeqa
In the context of the
type II string, $F_{g}^{II}$ only receives perturbative contributions
at genus $g$. In the heterotic context, $F_{g}^{het}$
is perturbatively determined at the tree and at the 
one loop level; in addition,
$F_g^{het}$ also receives non-perturbative corrections. 
For models with heterotic/type II duality
one expects that ${F}_g^{II}(t_i) = \alpha_g {F}_g^{het} (S,T_m)$,
where $\alpha_g$ denotes a normalisation constant.  The $t_i$ 
($i=1,\dots ,h$) 
denote the K\"ahler class moduli on the type II side, whereas $S$
and $T_m$ 
($m=1,\dots ,h-1$) denote the dilaton and the vector moduli on the heterotic
side.  

First consider the prepotential $F_0$ which determines the gauge
couplings.
The two prepotentials $F_0^{II}$ and $F_0^{het}$ should match up upon 
a suitable identification of the $t_i$ with $S$ and $T_m$.
On the type II side, the Yukawa couplings are given by
\cite{HKTY1}
\beqa
F^{II}_{klm} = {\cal F}^0_{klm} + \sum_{d_1,...,d_h} 
\frac{n^r_{d_1,...,d_h}d_k d_l d_m}{1-\prod_i^h q_i^{d_i}} 
\prod_{i=1}^h q_i^{d_i} ,
\eeqa
where $q_i = e^{- 2 \pi t_i}$.  The ${\cal F}^0_{klm}$ denote the
intersection numbers, whereas the $n^r_{d_1,...,d_h}$
denote the rational instanton numbers of genus zero.  These instanton numbers
are expected to be integer numbers.  
We will, in the following, work inside the K\"ahler cone
$\sigma(K)=\{\sum_i t_i J_i | t_i > 0\}$.
For points inside the
K\"{a}hler cone $\sigma(K)$, 
one has for the degrees $d_i$ that $d_i \geq 0$.

Integrating back yields that
\beqa
{F}_0^{II} = {\cal F}^0 - \frac{1}{(2 \pi)^3} \sum_{d_1,...,d_h}
n^r_{d_1,...,d_h} Li_3(\prod_{i=1}^hq_i^{d_i})
\label{ftype2}
\eeqa
up to a quadratic polynomial in the $t_i$.  
${\cal F}^0$ is cubic in the $t_i$.
Here one has used that $\partial_{t_k} \partial_{t_l} Li_3=
(-2\pi)^2 d_k d_l Li_1$, where $Li_1(x) = - \log(1-x)$.

In the following we will be focusing on a specific type IIA model, namely
the S-T-U model \cite{KaVa} based on the  Calabi-Yau space 
$WP_{1,1,2,8,12}(24)$ with $h=h_{11}=3$, $h_{21}=243$ and, hence, 
with $\chi=-480$.
Thus we have three K\"ahler moduli $t_1$, $t_2$, $t_3$ and instanton
numbers $n^r_{d_1,d_2,d_3}$.
The classical Yukawa couplings ${\cal F}^0_{klm}$ on the type II side are 
given by \cite{HKTY1}
\beqa
{\cal F}^0_{t_1t_1t_1}&=& 8 \;\;,\;\; 
{\cal F}^0_{t_1t_1t_2}= 2 \;\;,\;\;
{\cal F}^0_{t_1t_1t_3}= 4 \;\;,\;\; \nonumber\\
{\cal F}^0_{t_1t_2t_3}&=& 1 \;\;,\;\;
{\cal F}^0_{t_1t_3t_3}= 2 .
\eeqa
It follows that 
\beqa
{\cal F}^0 = \frac{4}{3} t_1^3 + t_1^2 t_2 + 2 t_1^2 t_3 + t_1 t_2 t_3 +
t_1 t_3^2 .
\label{fzero}
\eeqa
Some of the instanton numbers $n^r_{d_1,d_2,d_3}$ can be found in 
\cite{HKTY1}. 
When investigating  
the prepotential $F_0^{II}$
\cite{KLM},
two symmetries  become manifest, namely
\beqa 
t_1\rightarrow t_1+t_3,\; t_3\rightarrow -t_3
 \qquad {\rm for}\quad t_2=\infty ,\label{tu}
\eeqa
and 
\beqa
t_2\rightarrow -t_2,\; t_3\rightarrow t_2+t_3.\label{st}
\eeqa
These symmetries are true symmetries of $F_0^{II}$,  
since the world-sheet instanton numbers $n^r$
enjoy the remarkable properties \cite{KLM}
\beqa
n^r_{d_1,0,d_3}=n^r_{d_1,0,d_1-d_3} \qquad {\rm and}\quad 
n^r_{d_1,d_2,d_3}=n^r_{d_1,d_3-d_2,d_3}.\label{inssym}
\eeqa
Observe that ${\cal F}^0$ is completely invariant under the
symmetry (\ref{st}).

Next, consider the heterotic prepotential $F_0^{het}$.
$N=2$, $D=4$ heterotic strings can be constructed by compactifying
the ten-dimensional heterotic string on $T_2\times K_3$.
A generic compactification of the $E_8\times E_8$ heterotic string
on $K_3$, with equal $SU(2)$
instanton number in both $E_8$ factors, gives
rise to $D=6$ model with gauge group $E_7\times E_7$. For general
vev's of the massless hyper multiplets this gauge group is completely
broken, and one is left with 244 hyper multiplets and no massless
vector multiplets. Upon a $T_2$ compactification down to four dimensions,
one gets a model with 244 hypermultiplets and with
three vector multplets $S$, $T$ and $U$, 
where $S$ denotes the heterotic dilaton and $T,U$ denote the moduli of $T_2$.
This model is the heterotic dual
of the type IIA model considered above.
The heterotic prepotential has the following structure
\beqa
{F}_0^{het}=-STU+h^{(1)}(T,U)+{\cal F}^{\rm non-pert.}(e^{-S},T,U).
\label{fhetnp}
\eeqa
In the following we will consider the semiclassical limit $S\rightarrow\infty$,
i.e. ${\cal F}^{\rm non-pert}=0$, and 
we will concentrate on the 
one-loop corrected prepotential. 
The heterotic semiclassical prepotential 
\cite{WKLL,AntFerGaNaTay,CLM} has nontrivial
monodromy properties under the perturbative target space duality
symmetries $SL(2,{\bf Z})_T\times SL(2,{\bf Z})_U\times {\bf Z}_2^{T
\leftrightarrow U}$.
The singularities of the semiclassical prepotential at the lines/points
$T=U$, $T=U=1$ or $T=U=e^{i\pi /6}$ reflect the perturbative
gauge symmetry enhancement of $U(1)^2$ to $SU(2)\times U(1)$,
$SU(2)^2$ or $SU(3)$ respectively.
Derivatives of the semiclassical prepotential can be nicely expressed in
terms of automorphic functions of $T$ and $U$.
The semiclassical 
prepotential can be written in the following explicit form \cite{HarvMoo}
\beqa
{ F}_0^{het} &=& - STU + 
\frac{1}{384 \pi^2} \tilde{d}^{2,2}_{ABC} y^A y^B y^C\nonumber\\
  &-& \frac{1}{(2 \pi)^4} \sum_{k,l\geq 0} c_1(kl)
Li_3(e^{-2 \pi (k T + l U)})-{1\over (2\pi )^4}Li_3(e^{-2\pi (T-U)}),
\label{fhet}
\eeqa
where $y=(T,U)$ and where the constants $c_1(n)$ are 
related to the positive roots of a generalized Kac-Moody algebra.
These constants are determined by
\beqa
{E_4E_6\over\eta^{24}} = 
\sum_{n \ge -1} c_1(n) q^n &=& \frac{1}{q} - 240 - 141444 q -
8529280 q^2 - 238758390 q^3 \nonumber\\
&-&4303488384q^4 + \dots\label{fq}
\eeqa
The function ${F}_0^{het}$ has a branch locus at $T=U$.  ${F}_0^{het}$
given in (\ref{fhet}) is defined
in the fundamental Weyl chamber $T>U$.\footnote{It is meant here
that the real part of $T$ is larger than the
real part of $U$.}
The cubic coefficients $\tilde{d}^{2,2}_{ABC}$ will be  
determined below.  We have ignored a possible constant term as well as
a possible additional quadratic polynomial in $T$ and $U$.
The cubic terms cannot be uniquely fixed, since the prepotential
contains an ambiguity \cite{WKLL,AntFerGaNaTay}
which is a quadratic polynomial in the
period vector $(1,T,U,TU)$. Hence, the ambiguity is at most quartic in the
moduli and at most quadratic in $T$ and in $U$. It follows
that the third derivative in $T$ or in $U$ is unique; 
${\partial^2 h^{(1)}\over\partial T\partial U}$, however, is still ambiguous.
Specifically, in the chamber $T>U$, the cubic terms have the following
general form \cite{HarvMoo}
\beqa
\tilde{d}^{2,2}_{ABC} y^A y^B y^C
=  -32\pi \left( 3(1+\beta)T^2U + 3\alpha T U^2 + U^3 \right).
\label{cubichet}
\eeqa
The cubic term in $U$ is unique, whereas the  parameters $\alpha$ and $\beta$
correspond to the change induced by adding a quadratic polynomial
in $(1,T,U,TU)$.
As discussed in \cite{WKLL},  
it is convenient to introduce a dilaton field $S_{inv}$,
which is invariant under the perturbative $T$-duality transformations at the
one-loop level. It is defined as follows
\beqa
S_{inv}&=&S-{1\over 2}{\partial h^{(1)}(T,U)\over \partial T\partial U}
-{1\over 8\pi^2}\log(j(T)-j(U))\nonumber\\
&=&S+{1\over 4\pi}(1+\beta )T+{\alpha\over 4\pi}U
+ \frac{1}{8\pi^2} \sum_{k,l\geq 0} klc_1(kl)
Li_1(e^{-2 \pi (k T + l U)})\nonumber\\
&-&\frac{1}{8\pi^2} 
Li_1(e^{-2 \pi (T -  U)})
-{1\over 8\pi^2}\log(j(T)-j(U)).
\label{sinv}
\eeqa
In the decompactification 
limit 
to $D=5$ \cite{AFT}, obtained by sending $T,U\rightarrow\infty$ ($T>U$),
 the invariant dilaton $S_{inv}$ has
a particularly simple dependence on $T$ and $U$.  Namely, by 
using $\log j(T)
\rightarrow 2\pi T$, one obtains that 
\beqa
S_{inv}\rightarrow S_{inv}^\infty=S+{\beta\over 4\pi}T+{\alpha\over 4\pi}U.
\label{largetsinv}
\eeqa
Substituting $S_{inv}^\infty$ back into the heterotic prepotential 
(\ref{fhet}) yields that 
\beqa
{F}_0^{het} &=&  - S_{inv}^\infty TU -{1\over 12\pi}U^3 - {1\over 4\pi}T^2 U 
  - \frac{1}{(2 \pi)^4} \sum_{k,l \geq 0} c_1(kl)
Li_3(e^{-2 \pi (k T + l U)}) \nonumber\\
&-& \frac{1}{(2 \pi)^4} 
Li_3(e^{-2 \pi (T - U)}).
\label{fheta} 
\eeqa
Note that the ambiguity in $\alpha$ and $\beta$
is hidden away in 
$S_{inv}^\infty$.

Let us now compare the heterotic and the type II prepotentials and identify
the $t_i$ ($i=1,2,3$) with $S$, $T$ and $U$.
In the following,  we will actually match $ - 4 \pi { F}_0^{het}$
with ${F}_0^{II}$.
First compare the cubic terms in (\ref{ftype2}) and (\ref{fhet}).
By assuming that the $t_i$ and $S$, $T$ and $U$ are linearly related,
the following identification 
between the K\"ahler class moduli and the heterotic moduli is enforced
by the cubic terms
\beqa
t_1 &=& U  \nonumber\\
t_3 &= &T-U\nonumber\\
t_2 &=& 4\pi S_{inv}^\infty=\check{S} +\beta T + \alpha U 
\label{t2}
\eeqa
where $\check{S}=4 \pi S$. 
Recall that we are working inside the K\"ahler cone
$\sigma(K)=\{\sum_i t_i J_i | t_i > 0\}$.  
Now, in the heterotic
weak coupling limit one has that indeed $t_2 >0$. Demanding
$t_3 >0$ implies that one is choosing the chamber $T>U$ on the
heterotic side.  
The identification of $t_1$ and $t_3$ agrees, of course, with
the one of \cite{KLM}.
The identification of $4\pi S_{inv}^\infty$ with the K\"ahler
variable $t_2$ becomes very natural when performing the map to
the mirror Calabi-Yau compactification with complex structure
coordinates $x,y,z$. Here, since $y$ is invariant under
the CY monodromy group, $y$ should be identified \cite{KLM}
with $e^{- 8 \pi^2 S_{inv}}$.
Thus, equation (\ref{sinv}) provides the explicit mirror map;
for large $T,U$ the K\"ahler variable $q_2 = e^{- 2 \pi t_2}$ and the 
complex structure field $y$ completely agree.
%Furthermore, it is not difficult to explicitly show
%that the complex structure moduli $x$ and $z$ 
%agree
%with $q_1= e^{- 2 \pi t_1}$ and  $q_3 = e^{- 2 \pi t_3}$, respectively,
%in the limit $S,T,U\rightarrow\infty$ ($S>T>U$) \cite{KLM}.

Next, consider the exponential terms in the prepotential $F_0$.
In the heterotic weak coupling limit $S \rightarrow \infty$, one has that
$t_2 \rightarrow \infty$ and, hence, $q_2 = e^{-2 \pi t_2} \rightarrow 0$.
Then, (\ref{ftype2}) becomes 
\beqa
{F}_0^{II} = {\cal F}^0 - \frac{1}{(2 \pi)^3} \sum_{d_1,d_3}
n^r_{d_1,0,d_3} Li_3(q_1^{d_1}q_3^{d_3}) .
\label{ftype2weak}
\eeqa
Some of the instanton coefficients contained in 
(\ref{ftype2weak}) are
as follows\footnote{We are grateful to A. Klemm for providing us 
with a list of
instanton numbers for this model.} \cite{HKTY1}
\beqa
n^r_{d_1,0,0} &=&n^r_{d_1,0,d_1}= 480 = -2 (-240) \;\;,\;\; \nonumber\\
n^r_{0,0,1} &=& -2 \;\;\;,\;\;
n^r_{0,0,d_3} = 0 \;\;\;,\;\;\; d_3 = 2, \dots , 10 \;\;;\;\; \nonumber\\
n^r_{2,0,1} &=& 282888 = -2 (-141444) \;\;,\;\; \nonumber\\
n^r_{3,0,1} &=& n^r_{3,0,2} = 17058560 = -2 (-8529280) \;\;,\;\; \nonumber\\
n^r_{4,0,1} &=& 477516780 = -2 ( - 238758390) \;\;.\;\;
\eeqa
Note that the fact that $n^r_{d_1,0,0} = n^r_{d_1,0,d_1}$ 
is a reflection of the
$T \leftrightarrow U$ exchange symmetry.
Now rewriting $kT+lU = (l+k)U + k (T-U)= (l+k) t_1 + k t_3$
and matching ${ F}_0^{II} = - 4 \pi  {F}_0^{het}$ yields the
following identifications
\beqa
d_1 &=& k + l  \;\;,\;\;  d_3 = k \nonumber\\
n^r_{d_1,0,d_3} &=& n^r_{k+l,0,k} = -2 c_1(kl) .
\label{prepins}
\eeqa
Note that $d_3 = k \ge 0$ for points inside the K\"ahler cone.  Also, if 
$d_3=k=0$,
then $d_1 = l >0$.  On the other hand, if $d_3=k >0$, then 
$d_1 \ge 0$, that is $l \ge -k$.

Comparison of the instanton coefficients listed above
with the $c_1$-coefficients
ocurring in the $q$-expansion of $F(q)=\frac{E_4 E_6}{\eta^{24}}$ in 
equation (\ref{fq}),
shows that the relation (\ref{prepins}) is indeed satisfied.

Let us now determine the action of the symmetries (\ref{tu})
and (\ref{st}) on the heterotic variables.
Clearly the perturbative symmetry (\ref{tu}) corresponds
to the exchange $T\leftrightarrow U$ for $S\rightarrow\infty$.
The non-perturbative symmetry (\ref{st})
corresponds to 
\beqa 
S&\rightarrow & -(1+\beta )S-{\alpha (2+\beta)\over 4\pi}
U-{\beta(2+\beta)\over 4\pi}T\nonumber\\
T&\rightarrow &4\pi S+(1+\beta )T+\alpha U\nonumber\\
U&\rightarrow &U .
\label{hetst}
\eeqa
There is one very convenient choice for the parameters $\alpha$ and
$\beta$, in which the non-perturbative symmetry (\ref{hetst})
takes a very simple suggestive form. Namely, for $\alpha=0$ and 
$\beta=-1$, this transformation becomes
\beqa
4\pi S\leftrightarrow T,\label{stsim}
\eeqa
that is, it just describes the exchange of the heterotic dilaton $S$ with
the K\"ahler modulus $T$ of the two-dimensional torus.
This choice for $\alpha$ and $\beta$ is very reasonable,
since it is only in this case
that the real 
parts of $S$ and $T$ remain positive after the exchange 
(\ref{hetst}).
At the end of this paper, by considering \cite{AlFoIbaQue2}
some  six-dimensional one-loop gauge couplings, we will give some further 
arguments indicating that the choice
$\beta=-1$ is the physically correct one.
So, for the time being, we will set $\alpha=0$ and $\beta=-1$ 
and discuss a few issues related to  the exchange symmetry $4\pi S
\leftrightarrow T$.

The non-perturbative
exchange symmetry $4\pi S\leftrightarrow T$ 
is true for arbitrary $U$ in the chamber $S,T >U$.
As already discussed in detail in \cite{KLM}, 
at the fixed point $t_2=S_{inv}^\infty=0$ of 
this transformation, one has that $S=T>U$, the complex structure
field $y$ takes the value $y=1$, and the discriminant of the 
Calabi-Yau model vanishes. The locus
$S=T>U$ corresponds to a strong coupling
singularity with additional massless
states. In the model based on the Calabi-Yau space $WP_{1,1,2,8,12}(24)$,
a non-Abelian gauge symmetry enhancement with an equal number of
massless vector and hypermultiplets takes place at $S=T>U$,
such that the non-Abelian $\beta$-function vanishes \cite{KM,KaMoPle}.

On the other hand, the non 
perturbative exchange symmetry $4\pi S\leftrightarrow
T$ implies that for  $T\rightarrow\infty$ there is a `perturbative' 
$4\pi S\leftrightarrow U$ exchange symmetry. This symmetry is nothing
but the $T-S$ transformed perturbative symmetry (\ref{tu}).
Furthermore, for $T\rightarrow\infty$, there is a modular symmetry
$SL(2,{\bf Z})_S\times SL(2,{\bf Z})_U$ and the corresponding 
`perturbative' monodromy
matrices of the prepotential can be computed in a straightforward way. 
Hence, for $T\rightarrow\infty$, there is a `perturbative'
gauge symmetry enhancement of either $U(1)^2$ to $SU(2)\times U(1)$ or
to $SU(2)^2$ or to $SU(3)$ at the points $S=U$, $S=U=1$ or $S=U=e^{i\pi /6}$,
respectively, with no additional massless hyper multiplets \cite{CLM2}.

Let us now investigate the gravitational coupling $F_1$,
again first in the context of type II compactifications.
It can be expressed in terms of the K\"ahler moduli fields $t_i$ as 
the following instanton sum \cite{BerCecOogVa1}
\beqa
F_1^{II}=- i\sum_{i=1}^{h}
t_i c_2\cdot  J_i-{1\over\pi}\sum_n\biggl\lbrack
12n_{d_1,\dots ,d_h}^e\log(\tilde\eta(\prod_{i=1}^hq_i^{d_i}))+
n_{d_1,\dots ,d_h}^r\log(1-\prod_{i=1}^hq_i^{d_i})\biggr\rbrack.\label{ftop}
\eeqa
Here $\tilde\eta(q)=\prod_{m=1}^\infty(1-q^m)$, and
the $n_{d_1,\dots ,d_h}^e$
denote the elliptic genus one instanton numbers.
We will again specialize to the Calabi-Yau space $WP_{1,1,2,8,12}(24)$ with
$h=3$. Then the non-exponential piece, which dominates for large $t_i$,
reads \cite{KM}
\beqa
- i \sum_{i=1}^3t_i c_2\cdot J_i=92 t_1+24 t_2+48 t_3.\label{largef}
\eeqa
This expression is explicitly invariant under the non-perturbative
symmetry (\ref{st}).
Furthermore, by also explicitly
 checking some of the elliptic instanton numbers $n_{d_1,d_2,d_3}^e$,
one discovers that, just like in the case of $n^r$,  
\beqa
n^e_{d_1,0,d_3}=n^e_{d_1,0,d_1-d_3} \qquad {\rm and}\quad 
n^e_{d_1,d_2,d_3}=n^e_{d_1,d_3-d_2,d_3}.\label{inssyme}
\eeqa
It follows that $F_1^{II}$ is symmetric under the two exchange
symmetries (\ref{tu}) and (\ref{st}).

In the heterotic case
the holomorphic gravitational coupling at the one-loop level is given by
\beqa
F_{1}^{het} = 24 S_{inv} + \frac{b_{grav}}{8 \pi^2} \log \eta^{-2}(T)
\eta^{-2}(U) + \frac{2}{4 \pi^2} \log(j(T)-j(U)) .
\label{fgrav}
\eeqa
For the model we are discussing one has that $b_{grav}=48-\chi=528$.
Inserting $S_{inv}$ given in (\ref{sinv}) into
$F_{1}^{het}$ yields \cite{CLM2} 
\beqa
 F_{1}^{het} &=& 24 \left( S -
\frac{1}{768 \pi^2} \partial_T \partial_U
\left( \tilde{d}^{2,2}_{abc} y^a y^b y^c \right)
- \frac{1}{8 \pi^2} \log (j(T) - j(U)) \right. \nonumber\\
&+& \left. \frac{1}{8 \pi^2}
\sum_{k,l \geq 0} klc_1(kl)  L_{i_{1}}(e^{-2\pi (kT+lU)})  
- \frac{1}{8 \pi^2} L_{i_{1}}(e^{-2\pi (T-U)})  
\right) \nonumber\\
&+& \frac{b_{grav}}{8 \pi^2} \log \eta^{-2}(T)
\eta^{-2}(U) 
+ \frac{2}{4 \pi^2} \log(j(T)-j(U))  .
\label{fexp}
\eeqa
Let us now compare the heterotic and the type II gravitational
couplings.\footnote{It was already, to some extent, shown in
\cite{KaLouThei,Curio2,FeKhuMi}
that the
heterotic and type gravitational couplings agree.}
We will  match $4\pi F_{1}^{het}$ with $F_1^{II}$.
First take the decompactification limit to $D=5$,
i.e. the limit $T,U\rightarrow\infty$ ($T>U$). This eliminates all
instanton contributions, i.e. all exponential terms.
In the heterotic case we get in this limit
\beqa
F_{1}^{het}\rightarrow F_{1}^\infty=
24S_{inv}^\infty+{12\over \pi}T+{11\over\pi}U=
24S+{12+6\beta\over \pi}T+{11+6\alpha\over 4\pi}U.
\label{fgravinf}
\eeqa
By comparing this expression with the type II large $t_i$ 
limit given in (\ref{largef}), one finds that (\ref{largef}) and
(\ref{fgravinf}) match up precisely for 
the identification given in (\ref{t2}) between heterotic and type II
moduli.  When
choosing $\alpha=0$ and $\beta=-1$, it follows that
$F_{1}^\infty$ is symmetric under
the exchange $4\pi S\leftrightarrow T$.\footnote{In \cite{CLM2}
a different choice was made for these two parameters, namely 
$\alpha=-11/6$ and $\beta=-2$. Hence it follows 
that $F_{1}^\infty=24 S$.}
This symmetry implies that in the limit $T\rightarrow\infty$, $F_{1}^{het}$
can be written  \cite{CLM2}
in terms of $SL(2,{\bf Z})_S$ 
modular functions $j(4\pi S)$ and $\eta (4\pi S)$ 
by just replacing $4\pi S$ with $T$ in equation (\ref{fexp}).

Next, let us compare the exponential terms in $F_1^{II}$ and $F_{1}^{het}$.
In the type II case we 
have to consider the weak coupling limit $q_2\rightarrow 0$;
hence only the terms with the instanton numbers $n_{d_1,0,d_3}^{r,e}$
contribute to the sum. We will see that, when comparing with the heterotic
expression, one gets a very interesting relation between the
rational and elliptic instanton numbers for $d_2=0$.
In order to do 
this comparison, we have to recall that $Li_1(e^{-2\pi(kT+lU)})=
-\log(1-e^{-2\pi(kT+lU)})$.  The difference $j(T) -j(U)$ can
be written in the following useful form (in the chamber $T>U$) 
\cite{BORCH,HarvMoo}
\beqa
\log(j(T)-j(U))=2\pi T+\sum_{k,l}c(kl)\log(1-e^{-2\pi(kT+lU)}),
\label{diffj}
\eeqa
where the integers $k$ and $l$ can take the following values \cite{HarvMoo}:
either $k=1, l=-1$ or $k>0, l=0$ or $k=0, l>0$ or $k>0,l>0$.
The universal constants $c(n)$ are defines as follows:
\beqa
j(q)-744 =
\sum_{n=-1}^\infty c(n)q^n  &=&{1\over q}+196884q+21493760q^2+
864299970q^3 \nonumber\\
&+& 20245856256q^4 +\dots \label{cthree}
\eeqa
First consider the terms with $k=1,l=-1$ on the heterotic side.  
Matching the term $\log(1-e^{-2\pi{(T-U)}})$ contained in $4 \pi 
F^{het}_{1}$ with $F_1^{II}$ requires that
\beqa
10 c(-1) - 12 c_1 (-1) = 12 n^e_{0,0,1} + n^r_{0,0,1} .
\eeqa
This is indeed satisfied, since $c(-1) = c_1(-1) = 1$ and
$n^e_{0,0,1}=0, n^r_{0,0,1}=-2$.

Next, consider the terms in the sum with $k >0,
l=0$ (and analogously $k=0, l >0$). Since $c(0)=0$, only the term
${b_{grav}\over 8\pi^2}\log\eta^{-2}(T)$ contributes on the heterotic
side ($b_{grav}=528$). Matching $4\pi F_{1}^{het}$ with $F_1^{II}$
yields the following relation among the instanton numbers
($d_1=d_3=k$):
\beqa
12\sum_{i=1}^sn_{k_i,0,k_i}^e+n_{k,0,k}^r= b_{grav}=528.
\label{relf}
\eeqa
The $k_i$ ($i=1,\dots , s$) are the divisors of $k$ ($k_1=k$, $k_s=1$).
Using Klemm's list of 
explicit instanton numbers, we checked that this relation 
is indeed true up to $k=4$ ($n_{1,0,1}^e=4$, $n_{k,0,k}^e=0$ for $k>1$,
$n_{k,0,k}^r=-\chi=480$).

Finally, consider the case where $k>0,l>0$. By 
comparing the heterotic and type
II expressions we derive the following interesting
relation ($d_1=k+l$, $d_3=k$):
\beqa
12\sum_{i=1}^sn_{d_1^i,0,d_3^i}^e
&=&-n_{k+l,0,k}^r+10c(kl)+12klc_1(kl)=\nonumber\\ 
&=&10c(kl)+(12kl+2)c_1(kl).
\label{relt}
\eeqa
Here $s$ is the number of common divisors $m_i$
($i=1,\dots ,s$) of $d_1=k+l$ and $d_3=k$
with $d_1^i=d_1/m_i$ and $d_3^i=d_3/m_i$ (where $m_1=1$).
Again we can explicitly check the non-trivial relation (\ref{relt})
for the first few terms. For example, for $k=l=1$
one has
\beqa
n_{2,0,1}^e=-948 \qquad{\rm and}\qquad n_{2,0,1}^r=282888,
\eeqa
which, together with equations (\ref{fq}) and (\ref{cthree}), confirms the
above relation.
For $k=2$ and $l=1$ one finds that 
$12 n_{3,0,2}^e  +  n_{3,0,2}^r = 10 c(2) + 24 c_1(2)$
is indeed satisfied, since
\beqa
n_{3,0,2}^e=-568640 \qquad {\rm and}\qquad n_{3,0,2}^r=17058560.
\eeqa
And finally, for $k=l=2$ for instance, one finds that the relation
$12 \left(n_{4,0,2}^e  +  n_{2,0,1}^e\right) + 
n_{4,0,2}^r = 10 c(4) + 48 c_1(4)$ indeed holds due to
\beqa
n_{4,0,2}^e=-1059653772 \;\;\;,\;\;\;
n_{2,0,1}^e = -948
\qquad {\rm and}\qquad n_{4,0,2}^r=8606976768.
\eeqa
Now consider the relation $\biggl( \frac{E_6E_4}{\eta^{24}}
\biggr) '=-\frac{2\pi}{6}(\frac{E_2E_4E_6}{\eta^{24}}+2\frac{E_6^2}
{\eta^{24}}+3\frac{E_4^3}{\eta^{24}})$. 
{From} this we can, for $n >0$,
derive the useful equation 
$12nc_1(n)+10c(n)=-2\tilde{c}_1(n)$, where the $\tilde c_1(n)$ are defined
as follows
\beqa 
E_2\frac{E_6E_4}{\eta^{24}}(q) 
=  \sum_{n=-1}^{\infty}\tilde c_1(n)q^n=\frac{1}{q}-264 -135756 q -
5117440 q^2 + 
\dots \label{tilc}
\eeqa
It follows that one can 
rewrite  equation (\ref{relt}) as
\beqa
12\sum_{i=1}^sn_{d_1^i,0,d_3^i}^e+n_{d_1,0,d_3}^r
=10c(kl)+12klc_1(kl)=-2\tilde{c}_1(kl)\;,\; k>0, l>0.
\eeqa
This corresponds to the following integral\footnote{The precise
relation of this integral to $F_{1}^{het}$ was worked out in \cite{CLM2}.}
 of \cite{HarvMoo}
(with $b_{grav}=48-\chi=-2(c_1(0)-24)=-2\tilde{c}_1(0)$)
\beqa
\tilde I_{2,2}
&=&\frac{-1}{2}\int\limits_{\cal F} 
\frac{d^2\tau}{\tau_2}[-\frac{i}
{\eta^2}Tr_R J_0(-1)^{J_0}q^{L_0-22/24}\bar{q}^{\tilde{L}_0-9/24}
(E_2-\frac{3}{\pi\tau_2})-b_{grav}]\nonumber\\
&=&\frac{-1}{2}\int\limits_{\cal F} 
\frac{d^2\tau}{\tau_2}[-2Z_{2,2}\frac{E_4E_6}
{\eta^{24}}(E_2-\frac{3}{\pi\tau_2})-(-2\tilde{c}_1(0))]
\eeqa

Let us briefly summarize our results obtained so far. The type II prepotential
$F_0^{II}$
is determined by rational (genus 0) instanton numbers $n^r$. Comparison with
the semiclassical heterotic prepotential $F_0^{het}$
relates a subset of
the rational instanton numbers $n^r$ ($d_2=0$) to the coefficients of 
the modular function $E_4E_6\over \eta^{24}$ of modular weight -2.
The type II gravitational coupling $F_1^{II}$ depends, in addition, on the
elliptic (genus 1) instanton numbers $n^e$. A subset of those can be
expressed in terms of 
the coefficients of the modular functions $\frac{E_2E_4E_6}
{\eta^{24}}$ and $\frac{E_4 E_6}{\eta^{24}}$.
For the higher $F_g$ (cf. \cite{BCOV2}) we conjecture the following. 
Under modular transformations $T\rightarrow {1\over T}$
the $F_g$ transform at weak coupling as  
\beqa
F_g\rightarrow T^{2(g-1)}F_g,
\eeqa
i.e. $F_g$ has modular weight $2(g-1)$. Thus we are
tempted to conclude that the higher (genus $g$) instanton numbers
are determined by the coefficients of a modular form of modular weight
$2(g-1)$. Since the ring of modular functions, together with 
$\eta^{-24}$ is finite, only a finite number of different types of instanton
numbers seem to be independent. 

Note that such a fact is known for the case of a one dimensional Calabi-Yau
target space, that is an elliptic curve, where according to 
(also cf. \cite{D,R,K}) the $F_g$ are 
quasimodular forms of weight $6g-6$ for $g\geq 2$
, i.e. $F_g\in {\bf Q}[E_2,E_4,E_6]$.  Also note that 
one has $F_1=-\log \eta$ \cite{BerCecOogVa1}. 
This is here conjecturally extended to the 
elliptically fibered Calabi-Yau space considered above.

\section{Comments on other Calabi-Yau models}

At the end, let us 
briefly consider different
Calabi-Yau spaces and also comment on the relation
to the heterotic/heterotic duality in six dimensions 
\cite{DuMinWit,AlFoIbaQue2},
with the 6-dimensional heterotic string compactified on $K_3$.
(The decompactification limit from $D=4$ to $D=6$ is obtained 
by sending $T\rightarrow \infty$ with $U$ finite;
as discussed in \cite{Duff,DuMinWit}, the 
$D=6$ heterotic/heterotic duality becomes an
exchange symmetry of $S$ with $T$ in $D=4$.)
We will concentrate on three families of CY's 
with Hodge numbers (3,243), which are elliptic fibrations
over ${\bf F}_n$ with $n=0,1,2$ \cite{MoVa,SeiWit}. 
Being elliptic fibrations, they can be used to
compactify $F$-theory to six dimensions.
In  the $D=6$ heterotic string, 
the integer $n$ is related to the number $s$ of $SU(2)$ instantons
in one of the two $E_8$'s by $n=s-12$ \cite{MoVa,SeiWit}.
First consider the case of an elliptic fibration
over ${\bf F}_0$, corresponding to  the symmetric embedding of
the $SU(2)$ bundles with equal instanton numbers $s=s'=12$ into $E_8\times
E_8'$. This leads to a $D=6$ heterotic model with
gauge group $E_7\times E_7'$ with $\tilde v_\alpha
=\tilde v_\alpha '=0$ \cite{DuMinWit}. 
There are 510 hypermultiplets transforming as $4(56,1)+4(1,56)+62(1,1)$.
The heterotic/heterotic duality originates
from the existence of small instanton configurations \cite{Wit2}. 
The model is, however, not self-dual, since the non-perturbative gauge groups
appear in different points of the hyper multiplet moduli space than the
original gauge groups \cite{DuMinWit}. 
For generic vev's of of the hyper multiplets the gauge group is
completely broken and one is left with 244 hyper multiplets and no vector
multiplets.
Upon compactification to $D=4$ on $T_2$ one arrives
at the heterotic string with gauge group $U(1)^4$, 
which is the dual to the considered 
type II string on the CY $WP_{1,1,2,8,12}(24)$.
Semiclassically, at special points in
the hypermultiplet moduli space, this gauge group can be enhanced to
a non-Abelian gauge group, inherited from the $E_7\times E_7'$
with $N=2$ $\beta$-function coefficient $b_\alpha=12(1+{
\tilde v_\alpha\over
v_\alpha})=12$ \cite{AlFoIbaQue2}. 

Next, consider embedding the $SU(2)$ bundles in an asymmetric way into the
two $E_8$'s \cite{AlFoIbaQue1,AlFoIbaQue2}: $s=14$, $s'=10$. 
This corresponds to the elliptic
fibration over ${\bf F}_2$ \cite{MoVa,SeiWit}. 
Note that ${\bf F}_0$ and ${\bf F}_2$
are of the same parity (even N), so they are connected by deformation
\cite{MoVa}, as
we will discuss in the following.
Then, in this case, one has \cite{AlFoIbaQue2} a gauge group
$E_7\times E_7'$ with $\tilde v_\alpha=1/6$ and $\tilde v_\alpha '=-1/6$
and hyper multiplets transforming as
$5(56,1)+3(1,56)+62(1,1)$.
The second $E_7$ can be completely Higgsed away, leading to a 
$D=6$ 
heterotic model
with gauge group $E_7$ and hypermultiplets $5(56)+97(1)$.
As explained in \cite{AlFoIbaQue2},
this model also possesses a heterotic/heterotic
duality, however without involving non-perturbative small instanton
configurations. Hence in this sense, this model is really self-dual.
Just like in the case of the symmetric embedding, the gauge group $E_7$
is spontaneously broken for arbitrary vev's of the gauge non-singlet
hyper multiplets and one is again left with 244 hyper multiplets and
no vector multiplet. When compactifying on $T_2$ to $D=4$, one obtains
the same
heterotic string model with $U(1)^4$ gauge group as before.
For special values of the hyper multiplets a non-Abelian gauge group
is obtained, now however with 
$\beta$-function coefficient 
$b_\alpha=12(1+{\tilde v_\alpha\over v_\alpha})=24$ \cite{AlFoIbaQue2}.

In summary, the symmetric (12,12) model and the asymmetric (14,10) model
should be considered as being the same \cite{AlFoIbaQue2,MoVa},  since both 
are related by
the Higgsing and both lead
to the same heterotic string
in $D=4$. 
%This observation, however, creates a small puzzle
%concerning the non-perturbative dilaton moduli exchange symmetries in $D=4$. 
%Namely, since there are two types of 
%heterotic/heterotic duality transformations
%in $D=6$,  
%there should be  two different kinds of exchange symmetries also in $D=4$.
% Since the
%heterotic/heterotic duality transformation in the (14,10) model is precisely 
% self-dual it leads
%to the manifest exchange symmetry (\ref{stsim}), which can be understood
%by the properties (\ref{inssym}) and (\ref{inssyme}) of the instanton numbers.
%On the other hand,  the $D=6$ heterotic/heterotic
%duality in the symmetric (12,12) model could then correspond to the exchange
%of $S$ and $U$, a symmetry which is not visible in the chamber
%$S>T>U$.

As already mentioned, we  would like to provide a six-dimensional
argument for why  $\beta=-1$ is the physically correct choice
for one of the cubic parameters.
We will 
directly follow the discussion given in \cite{AlFoIbaQue2} 
and consider the one-loop gauge
coupling for the enhanced non-Abelian gauge groups that are inherited
from the six-dimensional gauge symmetries.
Specifically, the gauge kinetic function
is of the form \cite{WKLL,AlFoIbaQue2}
\beqa
f_\alpha=S_{inv}-{b_\alpha\over 8\pi^2}\log(\eta(T)\eta(U))^2.
\label{gaugec}
\eeqa
Using equation (\ref{largetsinv}) this then
becomes in the decompactification limit
$T\rightarrow\infty$ to $D=6$
\beqa
f_\alpha\rightarrow S+{1+\beta+{\tilde v_\alpha\over v_\alpha}\over 4\pi}T.
\label{flarge}
\eeqa
By comparing this expression 
with the six-dimensional gauge coupling \cite{Sagn}, it then follows that
$\beta = -1$.

Let us also make some remarks on the third model with Hodge
numbers (3,243), the (13,11)
embedding \cite{MoVa,SeiWit}. This is now elliptically fibered over 
${\bf F}_1$.  In going
to the Higgs branch \cite{SeiWit} one 
reaches the Calabi-Yau $WP_{1,1,1,6,9}(18)$
with $h^{1,1}=2$, $h^{2,1}=272$. Note that in the $D=6$ 
interpretation of F-theory
on this Calabi-Yau, this corresponds to loosing a tensor 
multiplet and gaining 29
hyper multiplets. 
Thus, in four dimensions, the
two vector multiplets correspond to $T$ and $U$.
No dilaton $S$ is present,
reflecting the fact that this CY is not a $K_3$
fibration, and no heterotic dual (at weak
coupling) exists. This CY is now elliptically fibered over ${\bf P}^2$ (the
exceptional curve of ${\bf F}_1$ was blown down)
\cite{SeiWit}. According to \cite{CFKM},
the rational instanton numbers $n_{j,0}^r$ of this CY are all equal to 
$540=-\chi _{CY}$.
Thus, compared to the CY $WP_{1,1,2,8,12}(24)$, the corresponding
modular form is now simply a constant. For $q_2=0$, the
Yukawa coupling $y_{111}$ of the
$WP_{1,1,1,6,9}(18)$ model is 
given by $E_4$ \cite{CFKM}. This can be nicely compared
with the following 
Yukawa coupling \cite{WKLL} of the CY $WP_{1,1,2,8,12}(24)$
model in the limit $T \rightarrow \infty$
\beqa
\partial_U^3 h^{(1)} \sim
\frac{E_4(U)}{j(T)-j(U)}\frac{E_4(T)E_6(T)}{\eta ^{24}(T)}\rightarrow E_4(U)
.
\eeqa
%thus possibly providing a modularity
%check in S (in the $U\rightarrow \infty$ sector), that is a connection
%between the $S\rightarrow \infty$ limit (the weak coupling computation)
%and the $S\rightarrow 0$ limit (the S-higgsed model).
%\footnote{This should be properly
%formulated in terms of $S^{inv}$.}
The elliptic instanton numbers in the $WP_{1,1,1,6,9}(18)$ model
satisfy the following relation:
$12\sum n_{j,0}^e+n_{j,0}^r=12\cdot 3+540$.  Here,  $n_{j,0}^e=3$ (versus 
$n_{j,0}^e=4$ in the $WP_{1,1,2,8,12}(24)$ model)
is determined by the elliptic fibration base with
$\chi({\bf P}^2)=3$ (versus  $\chi({\bf F}_1)=4$
in the $WP_{1,1,2,8,12}(24)$ model).  This
difference in the $n^e_{j,0}$ 
corresponds to the loss of one $h^{1,1}$ class (of the 
elliptic fibration base or equally well of the whole Calabi-Yau)
in the blowing down process. Accordingly, the  expression
$b_{grav}=48-\chi(CY)$ is modified.

It is instructive to consider the mirror map of this model.
Using the complex structure variable
 $Y_1=-\frac{1}{X_1}$ (cf. 
chapter 7.2 in \cite{CFKM}) the mirror map becomes
 $\frac{1}{Y_1(1-432Y_1)}=j(U)$ for $T\rightarrow\infty$,
 which corresponds to the
elliptic family $P_{1,2,3}(6)$ in \cite{KLM}.  
%This is in accordance 
% with the Calabi-Yau space $WP_{1,1,1,6,9}(18)$
%being fibered over ${\bf P}^2$ by this elliptic family.
This also corresponds to the $T\rightarrow \infty$
limit (at $S\rightarrow\infty$) of the $S-T-U$ Calabi-Yau model 
$WP_{1,1,2,8,12}(24)$, which
can also be considered to be elliptically fibered by the elliptic
family $P_{1,2,3}(6)$
over ${\bf F}_0$, when one considers \cite{AG,MoVa} the one  
nonpolynomial deformation of $WP_{1,1,2,8,12}(24)$, which deforms the base
${\bf F}_2$ to ${\bf F}_0$.

Let us close with the following remark. The $S-T$ exchange symmetry
is also present \cite{KLM} in the $S-T$ model based on the 
CY $WP_{1,1,2,2,6}(12)$ with Hodge numbers $h_{1,1}=2, h_{2,1}=128$.
This model, however, falls out of the class of the 
CY's considered above, since, even though being a $K_3$-fibration,
it does not correspond to an elliptic fibration. This model is
obtained \cite{KaVa}
by first performing a toroidal compactification to $D=8$ on a torus
with $T=U$ and enhanced $SU(2)$ gauge group,
and subsequently going down to $D=4$
by a $K_3$ compactification. Like in the case
of \cite{DuMinWit}, there is again a symmetric embedding
of the $SU(2)$ gauge bundle into $E_8\times E_8'\times SU(2)$:
$(s,s',s'')=(10,10,4)$.
The $S-T$ exchange symmetry, however, is not related to
a six-dimensional heterotic/heterotic duality or to $F$-theory on a CY.

\section{Acknowledgement}

We are grateful to A. Klemm for providing us with a list 
of instanton numbers for the CY model $WP_{1,1,2,8,12}(24)$.
We would also like to thank P. Berglund, A. Klemm, R. Minasian and especially
F. Quevedo and S.J.-Rey for fruitful discussions.

\end{document}